\documentclass[aps,twocolumn,showpacs,groupedaddress,amsmath,floatfix]{revtex4}
\usepackage{graphicx}
\usepackage{epsfig}
\usepackage{amsfonts}
\usepackage{amsmath,amssymb}

\newcommand{\be}{\begin{eqnarray}}
\newcommand{\ee}{\end{eqnarray}}
\newcommand{\non}{\nonumber\\}
 \newcommand{\ave}[1]{\langle {#1} \rangle}

 \newcommand{\GeV}{\hbox{GeV}}
 
 \newcommand{\Nch}{N_{\rm ch}}

\begin{document}


\title{Charge Transfer Fluctuations as a Signal for QGP}
\author{Lijun Shi}
\email[] {shil@physics.mcgill.ca}
\affiliation{Physics Department, McGill University,
Montr{\'e}al, Canada H3A 2T8}
\author{Sangyong Jeon}
\email[] {jeon@physics.mcgill.ca}
\affiliation{Physics Department, McGill University,
Montr{\'e}al, Canada H3A 2T8}
\affiliation{RIKEN-BNL Research Center, Upton NY 11973, USA}

\date{\today}

\begin{abstract}
In this work, the charge transfer fluctuation which was previously
used for $pp$ collisions is proposed for relativistic heavy-ion
collisions as a QGP probe.
We propose the appearance of a local minimum at midrapidity for
the charge transfer fluctuation as a signal for a QGP.
Within a two-component neutral cluster model, we
demonstrate
that the charge transfer fluctuation can detect the presence of a QGP
as well as the size of the QGP in the rapidity space.
We also show that the forward-backward correlation of multiplicity
can be a similarly good measure of the presence of a QGP.
Further, we show that the previously proposed net charge fluctuation
is sensitive to the existence of the second phase only if the QGP
phase occupies a large portion of the available rapidity
space.
\end{abstract}

\maketitle

\section{Introduction \label{sect::intro}}
Active researches in relativistic energy heavy-ion collisions have given
us
much information about the hot matter produced in such collisions. Much
attention has been directed to the question of whether a deconfined
quark-gluon
plasma (QGP) phase has been formed. The experimental
studies do suggest that
strongly interacting dense matter was formed during the early stage of
reaction,
and the energy density of such matter is very high (see
\cite{Adcox:2004mh,Back:2004je,Adams:2005dq,Arsene:2004fa,Shuryak:2004cy}
and
references therein).
Most theoretical models for such hot and dense matter explicitly invoke
quark
and gluon degrees of freedom in the elementary  processes
\cite{Blaizot:2001nr,Iancu:2000hn,Ferreiro:2001qy,Blaizot:1999xk,
Wang:1991ht,Wang:1991us,Gyulassy:1994ew,Wang:1996yf,Sorge:1993nv,
Sorge:1995dp,Bass:1998ca,Bleicher:1999xi,Gale:1987ki,Lin:2002gc}.
One way to detect the presence of a QGP is then to measure the changes in
the
fluctuations and correlations which could
originate from the new phase of matter.

In this work, we propose charge transfer fluctuations
as a signal of
the presence of a QGP as well as the measure of the (longitudinal)
size of the QGP.
The charge transfer fluctuation for elementary collisions
was originally proposed by Quigg and Thomas \cite{Quigg:1973wy}
where they considered
a flat charged particle distribution $dN_{\rm ch}/dy$. This idea
was later extended to smooth distributions
by Chao and Quigg\cite{Chao:1973jk}.

The central result of Refs.\cite{Quigg:1973wy,Chao:1973jk}
is the relationship between
the single particle distribution
function $dN_{\rm ch}/dy$ and the charge transfer fluctuation:
\begin{eqnarray}
   D_u(y) = \kappa \frac{dN_{\rm ch}}{dy}
  \label{eq:ChTr:Chao-Quigg-smooth}
\end{eqnarray}
where $\kappa$ is a constant and
  \begin{eqnarray}
  D_u(y) \equiv \langle u(y)^2 \rangle - \langle u(y)\rangle ^2.
  \label{eq:ChTr:chargetransfluc-def}
  \end{eqnarray}
  is the charge transfer fluctuation.
The charge transfer $u(y)$
is defined by the forward-backward charge difference:
  \begin{eqnarray}
  u(y) &=& \left[ Q_F(y)-Q_B(y) \right] /2  \,,
  \label{eq:ChTr:fbcharge-asymmetry}
  \end{eqnarray}
where $Q_F(y)$ is the net charge in the rapidity region
forward of $y$ and $Q_B(y)$ is the net charge in the rapidity region
backward of $y$.
The fluctuation $D_u(y)$ is then a measure of the correlation between the
charges in the forward and the backward regions separated by $y$.

\begin{figure}[t]
\centerline{ \includegraphics[scale=0.40]{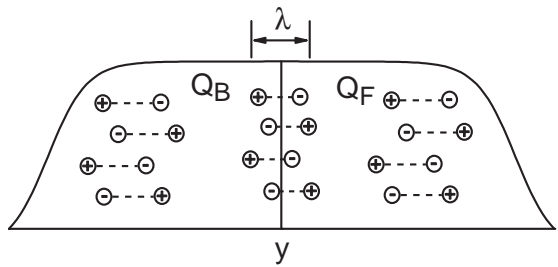}}
\caption{A schematic illustration of
the charge transfer fluctuations in the rapidity space.
Only the pairs within $\lambda/2$ of $y$ can
contribute to the charge transfer fluctuation $D_u(y)$.
Here $\lambda$ is the rapidity correlation length, or the rapidity
distance of the decay particles from a single cluster.
If $\lambda$ is a function of $y$, then $D_u(y)$ also changes
with $y$.}
\label{fig:concept0}
\end{figure}

The importance of the relationship (\ref{eq:ChTr:Chao-Quigg-smooth})
lies in the fact
that $\kappa$ is in fact directly proportional to the {\em local}
unlike-sign charge correlation length.
Heuristically, this can be explained in the following way.
Suppose all final particles originate from neutral clusters and
each cluster produces
one positively charged particle and one negatively charged particle.
Then the only way $u(y)$ can deviate from zero is when one charged
particle from a cluster ends up in the forward region while the other
ends up in the backward region as
illustrated in Fig.\ref{fig:concept0}.
For each one of these split pairs, the charge transfer $u(y)$ undergoes
a 1-D random walk with a step size 1.
Therefore, the charge transfer fluctuation
$D_u(y)$ should be proportional to
the number of split pairs, or equivalently the number of random steps
taken.

If $\lambda$ is
the typical rapidity difference between the two decay particles from a
single cluster,
then only the clusters within the rapidity
interval $(y-\lambda/2, y+\lambda/2)$ can contribute to $D_u(y)$ as
illustrated in Fig.\ref{fig:concept0}.
The number of such clusters is then
$\lambda dN_{\rm clstr}/dy$ where $dN_{\rm clstr}/dy$ is the
density of the clusters at $y$.
Since the final particle spectrum $dN_{\rm ch}/dy$ should be proportional
to
$dN_{\rm clstr}/dy$,
we have Eq.(\ref{eq:ChTr:Chao-Quigg-smooth}) with
$\kappa\propto \lambda$.
Hence the ratio
$\kappa = D_u(y)/(dN_{\rm ch}/dy)$
is a measure of the {\em local} environment near $y$:
If $\lambda$ is a function of $y$, then $\kappa(y)$
should also change accordingly.

The fact that $\kappa$ is constant in elementary collisions
indicates that in such collisions
the correlation length is constant throughout the entire rapidity
range (see \cite{Kafka:1975cz,Chao:1973jk} and references therein).
However, if a QGP is produced in the central region of the relativistic
heavy ion collisions, we can expect the local charge correlation length
$\gamma(y)$ increases as $y$ moves away from central rapidity.
This is because the charge correlation length in a QGP is expected to be
much smaller than that in a hadronic phase~\cite{Bass:2000az,Koch:2001zn}.
In this case, the ratio
  \begin{eqnarray}
  \kappa(y) = {D_u(y)\over {dN_{\rm ch}/dy}} \,,
  \end{eqnarray}
will vary from a smaller value to a larger value as one goes away from the
central region toward the forward region.

There have been many studies of the fluctuations and correlations
in heavy ion collisions
\cite{Shuryak:1997yj,Stephanov:1999zu,
Mrowczynski:1999un,Korus:2001fv,
Stodolsky:1995ds,Asakawa:2000wh,
Jeon:2000wg,Mitchell:2004xz,Lindenbaum:2001gw,Adcox:2002mm,
Adams:2003st,Pruneau:2004nc,Nystrand:2003ed,Bass:2000az,
Jeon:2001ue,Cheng:2004zy,Adams:2003kg,Tonjes:2002us,Westfall:2004cq}.
Most of these studies concentrate on {\em global} information and do not
address possible spatial inhomogeneity of the created matter in
relativistic
heavy-ion collisions. For instance,
if the QGP phase is confined to a small rapidity region, the net charge
fluctuation measures proposed in Refs.\cite{Asakawa:2000wh,Jeon:2000wg}
may not be very sensitive to the presence of the QGP.
Hence negative results
from experiments~\cite{Adcox:2002mm,Adams:2003st,Mitchell:2004xz}.
do not necessarily exclude the formation of a QGP.

Our expectation that the central rapidity region in the
heavy ion collisions is
mostly QGP originates from Bjorken's seminal
work~\cite{Bjorken:1983AA}. In that paper it was assumed
that the expanding QGP
evolves in a boost-invariant manner.  Such an assumption
naturally leads to the
expectation that the central plateau in the rapidity
spectrum is a manifestation
of a boost-invariant QGP. However, recent RHIC results cast some doubts on
boost-invariant scenario in the central rapidity region: Although the
charged particle distributions as a function of
{\it pseudo-rapidity} shows a
central plateau \cite{Back:2001xy,Back:2004bq}, the recent
{\em rapidity} spectrum of charged particles from the BRAHMS group
is consistent with a gaussian following the Landau
picture~\cite{Bearden:2004yx} although a plateau within
$-1 < y < 1$ cannot be ruled out~\cite{Adams:2003yh}.
The elliptic flow spectrum from the PHOBOS
group~\cite{Back:2004mh} shows no discernible plateau
at all as a function of the pseudo-rapidity.
Thus, a simple boost invariance scenario
in a large range of rapidity space as originally envisioned by Bjorken
\cite{Bjorken:1983AA} may not be valid. If the QGP phase is produced in
the
relativistic heavy-ion collisions,
a pure phase may very well be confined to a very limited rapidity range.
It is, therefore, important to have an observable that is sensitive to the
{\em local} presence of a QGP.
The charge transfer fluctuation is such a local measure of a phase change.

Of course, as emphasized in Ref.~\cite{Pruneau:2002yf},
a particular type of
fluctuations is just one particular aspect of the underlying correlations.
Usefulness of each
type of fluctuations then depends on the sensitivity of the chosen
fluctuation to an interesting aspect of the correlation. For charge
transfer fluctuations,
that aspect is the size of the local charge correlation length.
Hence if the QGP
phase is spatially confined to a narrow region around the midrapidity,
the charge transfer fluctuations can signal its presence and also
can yield information about the size.

In this study, we propose
the appearance of a clear minimum at midrapidity for the ratio $\kappa(y)$
as a signal for the existence of two different phases.
The slope and the size of
the dip around midrapidity can then reveal the size of the new phase
(presumably a QGP).
These features should disappear as the energy is lowered or the
collisions become more peripheral where a QGP is not expected to form.

In the following, we use a single component neutral cluster model and a
two component neutral cluster model to study the purely hadronic case
and the mixed phase case.  However, the fact that the charge transfer
fluctuation is a useful measure of the {\em local} correlation length is
independent of our particular choice of models.  Hence we expect that the
general conclusions drawn in this study should be valid even within more
sophisticated models as well as in real experimental
situations.
A case study using cascade models
with an embedded QGP component is under way.

We note here that
most of the discussions in this study are in terms of the rapidity $y$.
However, the validity of our results does not depend very much on whether
rapidity $y$ is used or the pseudo-rapidity $\eta$ is used.
We also note here that
the argument given here applies with very little change to any conserved
charges such as the baryon number.

The rest of this
paper is organized as follows: In the next section, we consider the
basic phenomenology of the charge transfer fluctuations. In
Sect.\ref{sect::one-comp}, we consider the net
charge fluctuations and the charge transfer fluctuations in a single
component
model. In Sect.\ref{sect::two-comp}, we present our main results on a two
component model.
It is proposed that the presence of a
rising segment of the charge transfer fluctuation as a function of
rapidity
can be used as a QGP signal.
We also show that the charge transfer fluctuation can reveal the size of
the QGP. A summary is given in Sect.\ref{sect::summary}.

\section{Charge Transfer Fluctuations
\label{sect::charge-transfer}}

The charge
transfer is defined
in Eq.(\ref{eq:ChTr:fbcharge-asymmetry}).
The charge transfer fluctuation is defined in
Eq.(\ref{eq:ChTr:chargetransfluc-def}).
Originally Quigg and Thomas \cite{Quigg:1973wy},
considering a flat $dN_{\rm ch}/dy$,
argued that if all hadrons originated from neutral clusters, then the
following relation should hold:
  \begin{eqnarray}
  {D_u(0)} = \frac{4 \lambda}{3}\, \frac{ N_{\rm ch}}{Y_{\rm max}}
  \label{eq:ChTr:Quigg-Thomas-flat}
  \end{eqnarray}
where $\lambda$ is the rapidity correlation length of unlike-sign
$(+-)$ pairs originating from a single neutral cluster.
$N_{\rm ch}$ is the total multiplicity of the produced charged particles
and
$Y_{\rm max}/2$ is the beam rapidity in the CM frame.

\begin{figure}[t]
\center
\includegraphics[scale=0.30,angle=90]{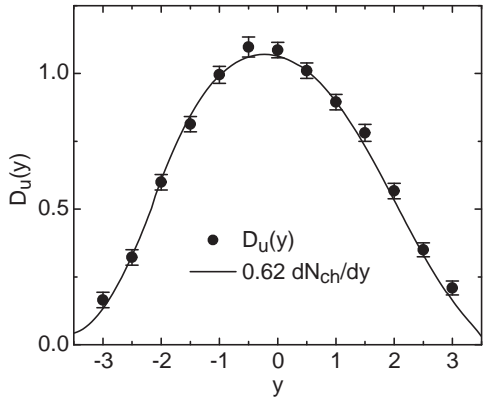}
\caption{
The charge transfer fluctuation results from $pp$
collisions at the beam energy of $205\,\GeV$ is shown as a
function of rapidity. The line is the charge yield profile
measured in the same
reaction scaled by a factor of $0.62$. This is a
re-plot the results reported in
\cite{Kafka:1975cz}.
\label{fig:pp205} }
\end{figure}

Later, this was extended by
Chao and Quigg to smooth charged particle distributions~\cite{Chao:1973jk}
to yield Eq.(\ref{eq:ChTr:Chao-Quigg-smooth}), $D_u(y)=\kappa d\Nch/dy$,
with $\kappa \propto \lambda$. The experimental results on $pp$ and $K^-p$
collisions show that this relationship is remarkably good with $\kappa$
ranging from $0.62$ to $0.85$
(see \cite{Kafka:1975cz,Chao:1973jk} and
references therein). We have re-plotted $205$ GeV $pp$ collision results
from Ref.\cite{Kafka:1975cz} in Fig.\ref{fig:pp205}.
Here the proportionality constant $\kappa$ is approximately $0.62$.
As we shall see later in this section, the proportionality of the charge
transfer fluctuation to the charged particle spectrum
is a strong argument for correlated charge pairs instead
of uncorrelated charged particles.

The charge correlation length $\lambda$ measures the
rapidity correlations
between unlike-sign charges.
(This quantity also plays
a central role in two of the
proposed QGP signals, namely the net charge
fluctuations and the balance function.)
To illustrate the relationship between
the proportionality constant $\kappa$ and
the unlike-sign charge correlation length $\lambda$ in
Eq.(\ref{eq:ChTr:Chao-Quigg-smooth}), we consider a simple `$\rho$'
gas model (see e.g. \cite{Jeon:1999gr}).

In this  model, each `$\rho^\pm$' is assumed to
decay to a $\pi^0 \pi^\pm$ pair and each `$\rho^0$'
is assumed to decay to a $\pi^+\pi^-$
pair. This is similar to the `$\rho$' and `$\omega$' models used for $pp$
collisions \cite{Chao:1973jk,Berger:1973nc}.
One should not, of course,
regard these $\rho$'s as physical $\rho$-mesons.
These are just convenient names for charged and neutral clusters.
In particular, they are not isospin triplets.

Consider a set of events
where $M_0$ number of  $\rho^0$'s and
$M_+$  and $M_-$ number of
$\rho^+$'s and $\rho^-$'s are produced.
The full joint probability for the rapidity
of the charged pions for this set is given by:
  \begin{eqnarray}
  \rho(\{y_a\})=
  \prod_{i=1}^{M_{+}}g(y_i)
  \prod_{j=1}^{M_{-}}g(y_j)
  \prod_{k=1}^{M_0}f_0(y_k^+, y_k^-) \,,
  \label{eq:ChTr:ResGasMod}
  \end{eqnarray}
where $f_0(y^+, y^-)$ is the probability for
the two decay products of a
neutral cluster to have the
rapidities $y^+$ and $y^-$ and $g(y)$ is
the single particle distribution
function for the charged particles originating
from the charged $\rho$'s.
Averaging over the distributions of $M_0, M_+, M_-$ with
the condition $Q = M_+ - M_- = {\rm constant}$,
it is not hard to show that
  \begin{eqnarray}
  D_u(y) &=&
  2\left\langle M_0 \right\rangle
  \int^y_{-\infty} dy'\int_y^\infty dy'' \, f_0(y', y'')
  \nonumber\\
   & & +
  \left\langle M_{\rm ch} \right\rangle
  \int^y_{-\infty} dy' g(y') \int_y^\infty dy'' \, g(y'') \, .
  \label{eq:ChTr:ChTrFl-dist-func}
  \end{eqnarray}
  where $M_{\rm ch} = M_+ + M_-$.
In arriving at the above result, we neglected a contribution
that is on the order of
$\langle u(y) \rangle^2 / N_{\rm ch}\sim Q^2/N_{\rm ch}$ where
$N_{\rm ch} = 2\ave{M_0}{+}\ave{M_{\rm ch}}$. This should be
small compared to the terms in Eq.(\ref{eq:ChTr:ChTrFl-dist-func}) when
$Q \ll N_{\rm ch}$.  See Appendix B for details.

The single particle rapidity distribution is given by
  \begin{eqnarray}
  \frac{dN_{\rm ch}}{dy} =  2\left\langle M_0 \right\rangle h(y)
   +\left\langle M_{\rm ch}  \right\rangle g(y) \,,
  \label{eq:ChTr:ChDens-dist-func}
  \end{eqnarray}
where $h(y)  = \int_{-\infty}^\infty dy'\, f_0(y', y)$.

The Thomas-Chao-Quigg relationship, $D_u(y) = \kappa dN_{\rm ch}/dy$,
can be
solved explicitly in two extreme cases when either $M_0 = 0$ or $M_{\rm
ch}=0$.  When $M_0 = 0$, we have
  \begin{eqnarray}
  \int^y_{-\infty} dy' g(y') \int_y^\infty dy'' \, g(y'')
  =
  \kappa g(y)
  \end{eqnarray}
and the solution is given by
  \begin{eqnarray}
   g(y) = \frac{1}{4\kappa} {\rm sech}^2(\frac{y}{2\kappa})
   \label{eq:ChTr:ChTrFl-OneSolve-g}
   \end{eqnarray}
   which can be easily verified using
$\hbox{sech}^2 x = d\tanh x/dx = 1-\tanh^2x$.
With $\kappa = O(1)$, this form alone (basically the
modified P\"oschl-Teller potential) is much too sharp to
describe a realistic $dN_{\rm ch}/dy$.
Furthermore, it
has no room for energy dependence once $\kappa$ is fixed. This is in
contradiction with the $dN_{\rm ch}/dy$ spectrum in elementary
collisions which shows no
prominent central peak of a fixed width.
Hence the $M_0 = 0$ scenario can be excluded.

In the $M_{\rm ch} = 0$ limit,
the Thomas-Chao-Quigg relationship is
\begin{eqnarray}
  \int^y_{-\infty} dy'\int_y^\infty dy'' \, f_0(y', y'')
  =
  \kappa \int_{-\infty}^\infty dy'\, f_0(y', y)
  \label{eq:ChTr:onlyf0}
\end{eqnarray}
To solve for $f_0(y,y')$, we make an ansatz
  \begin{eqnarray}
  f_0(y, y') = R(y_{\rm rel})\, F(Y) \,,
  \label{eq:ChTr:f0-ansatz}
  \end{eqnarray}
where $y_{\rm rel} = y-y'$ and $Y = (y+y')/2$. The normalization
conditions
for
$R$ and $F$ are $\int_{-\infty}^\infty dy\, R(y) = \int_{-\infty}^\infty
dy\,
F(y) = 1$.
Equation E(\ref{eq:ChTr:onlyf0}) can then be
solved by making change of variables
to $y_{\rm rel}$ and $Y$.
The solution is
  \begin{eqnarray}
  f_0(y, y') = \frac{1}{4\kappa}
  \exp{\left(-\frac{|y-y'|}{2\kappa}\right)}\,
  F\left(\frac{y+y'}{2}\right)
  \label{eq:ChTr:f0-generalform}
  \end{eqnarray}
where the only restriction on $F$ is that the integrals in
Eq.(\ref{eq:ChTr:onlyf0}) converge and it reproduces the experimental
$dN_{\rm ch}/dy$.
For details, see Appendix~\ref{app:solution}.
This form of correlation function is very reasonable as this is
nothing but the distribution function of the decay products
of a cluster whose rapidity is distributed according to $F(Y)$.

For small enough $\kappa$, we should have
  \begin{eqnarray}
  F(y) \approx \frac{1}{N_{\rm ch}} \frac{dN_{\rm ch}}{dy}
  \label{eq:ChTr:Fy-narrow-kappa}
  \end{eqnarray}
where $N_{\rm ch} = 2 \left\langle M_0 \right\rangle $
is the total charge multiplicity.
In this solution, it is clear that $\kappa$ is directly related to
the correlation length $\lambda$ in the relative rapidity space
of the pair $y-y'$ as
  \begin{eqnarray}
  \lambda = 2 \, \kappa
  \end{eqnarray}

Taking the correlation $\kappa \sim 1$, it is easy to see that the charged
particle spectrum
will be a smooth distribution with typical variation in rapidity of
$2\kappa
\sim 2$. This is in good agreement with the $pp$ collision results in
Fig.\ref{fig:pp205}. Further, the absence of a narrow peak in the central
region
also excludes significant contribution from the uncorrelated charged
particles.

If we have a finite observational window $(-y_{\rm o}, y_{\rm o})$,
the charge
transfer fluctuations are given by:
  \begin{eqnarray}
 \bar{D}_u(y) &=& \frac{\langle M_0 \rangle}{2} \Bigg[
  \int_{-y_{\rm o}}^{y_{\rm o}} dy'\int_{-\infty}^{-y_{\rm o}} dy''
  \, f_0(y',y'')
 \nonumber \\
 &&  +
  \int_{-y_{\rm o}}^{y_{\rm o}} dy' \int_{y_{\rm o}}^{\infty} dy''
  \, f_0(y',y'')
 \nonumber \\
 & &  +
  4\int^y_{-y_{\rm o}} dy'\int_y^{y_{\rm o}} dy'' \, f_0(y', y'')  \Bigg]
  \label{eq:ChTr:Duy-general}
  \end{eqnarray}
where the bar in $\bar{D}_u$ indicates that this quantity is measured in a
limited window.
With a moderate $y_{\rm o}$,
the Thomas-Chao-Quigg relationship Eq.(\ref{eq:ChTr:Chao-Quigg-smooth})
no longer holds even if the unrestricted charge transfer fluctuation
satisfies it.  However $\bar{D}_u(y)$ is still sensitive to the charge
correlation length.
To have an idea how $\bar{D}_u(y)$ behaves, we can use a flat $F(Y)$
following Thomas and Quigg.
If $\kappa$ is much smaller than the size of the total rapidity
interval $Y_{\rm max}$
and $y_{\rm o}$ is not too close to $Y_{\rm max}/2$, we get
  \begin{eqnarray}
  \frac{\bar{D}_u(y)}{ \left\langle N_{\rm ch} \right\rangle } =
   \frac{\kappa}{ 2 y_{\rm o}} \left[{\frac{3}{2}}
   + \frac{1}{2}e^{-y_{\rm o}/\kappa}
   - 2e^{-y_{\rm o}/2\kappa}\,\cosh\left(\frac{y}{2\kappa}\right)\right]
  \label{eq:ChTr:Fy-flat-solveDuy}
  \end{eqnarray}
If we have a large $y_{\rm o}/\kappa$,
this becomes constant and we get back
to the original Thomas-Quigg relationship
Eq.(\ref{eq:ChTr:Quigg-Thomas-flat}).  With a finite $y_{\rm o}$, this is
a
monotonically decreasing function of $y > 0$ with the maximum at $y=0$.

The pseudo-rapidity distribution measured by RHIC experiments does show a
plateau within $-2 < \eta < 2$ and the rapidity distribution shows a
similar
plateau within $-1 < y < 1$.
Hence, if the correlation length ($\kappa$)
remains constant throughout the plateau, one would expect that
$\bar{D}_u(y)$ measured within the plateau should also have
a maximum at $y=0$.
On the other hand, if one observes that $\bar{D}_u(y)$
has a {\em minimum} at $y=0$, then it is a signal that
quite a different system (a QGP) is
created near the central rapidity with a much smaller rapidity
correlation length than the rest of the system.

\begin{figure}[t]
\includegraphics[scale=0.30,angle=90]{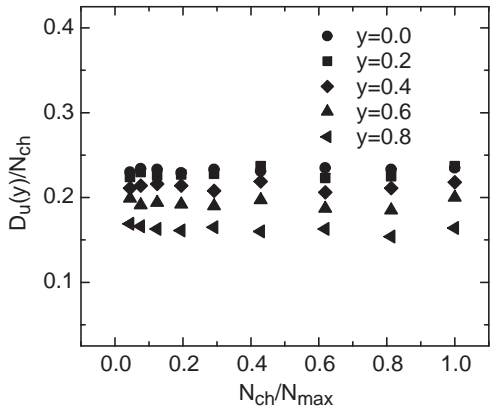}
\caption{
The ratio of the Charge transfer fluctuations to the number of charged
particles $\bar{D}_u(y)/N_{\rm ch}$ is shown as a function of centrality
$N_{\rm ch}/N_{\rm max}$.
The observation window is fixed at $(-1.0,1.0)$. These results are from a
HIJING
calculation for $\sqrt{s}=200$ GeV Au-Au collisions. Parameters for the
HIJING
calculations are $dE/dx=2$ GeV/fm, $p_{minijet}=2$ GeV with the shadow and
the
quench flags on.
The most central bin has the top $5\,\%$ of the events.
All other bins have the size of $10\,\%$.
Total of $50,000$ HIJING events are analyzed.
\label{fig:hijing}}
\end{figure}

To examine the relationship between the charge transfer fluctuations and
the
charge correlation length, we must make sure that other effects such as
the
impact parameter fluctuations and the hadronic correlations do not mimic
the
effect we seek. To study the non-QGP effects, we have analyzed $50,000$
HIJING
events \cite{ToporPop:2002gf}. The results are shown in
Fig.\ref{fig:hijing}.
Each point in the figure represents $5\%$ bin in the centrality measured
by
the
number of charged particles within $-1 < y < 1$. It is quite obvious from
this
figure that the charge transfer fluctuations do not vary with centrality.
It is also obvious that $\bar{D}_u(y)$ in this case is a decreasing
function
of $y$.

\begin{figure}[t]
\includegraphics[scale=0.4]{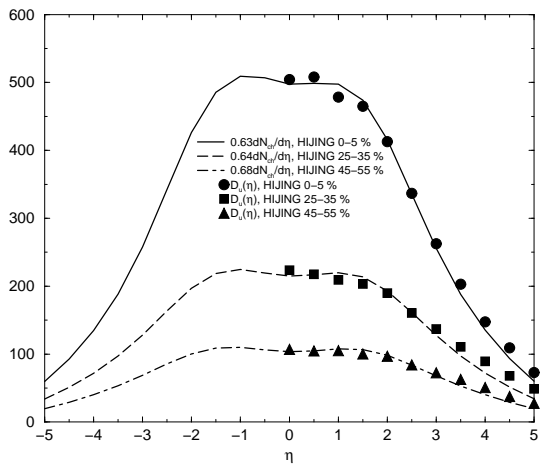}
\caption{
The results of analyzing simulated Au-Au events using HIJING.
The lines are scaled pseudo-rapidity spectra for centrality classes
$0-5\,\%$, $25-35\,\%$ and $45-55\%$ and the symbols are charge transfer
flucuations for each class.
\label{fig:hijing_full}}
\end{figure}

It is interesting to compare Eq.(\ref{eq:ChTr:Fy-flat-solveDuy}) with the
results from HIJING. For this, we used the top $15\%$ central results in
Fig.\ref{fig:hijing} and fitted them with
Eq.(\ref{eq:ChTr:Fy-flat-solveDuy}).
The best fit gives $\kappa = 0.72$ though the results from
Eq.(\ref{eq:ChTr:Fy-flat-solveDuy}) are slightly flatter than the HIJING
results.
This discrepancy in shape
is not unexpected because the HIJING $dN_{\rm ch}/dy$
is not well approximated by the
flat $F(Y)$.
The full pseudo-rapidity space analysis of 50,000 HIJING minimum bias
events are shown
in Fig.\ref{fig:hijing_full} for 3 different centralities.
The Thomas-Chao-Quigg relationship works quite well within this model for
all centrality classes.
In the rapidity region of $y=0 \sim 3$,
the value of $\kappa$ is in the range of $0.63 \sim 0.68$.
This is consistent with the
experimental $pp$ result.

In the next two sections, we test our idea against two scenarios for heavy
ion collisions.
In the first scenario,
the created system consists of a single species of neutral clusters which
may be taken as hadronic clusters.  In the second scenario, the created
system contains a second component with a much smaller correlation length.

\section{Single Component Model\label{sect::one-comp}}

In this section, we consider the charge transfer fluctuations in a
system which consists of only a single species of neutral clusters
(presumably a hadronic matter).

The Thomas-Chao-Quigg relationship
was used to justify the use of the neutral cluster model in $pp$
collisions in the last section.
For heavy ion collisions, as far as we know there has been no
experimental investigation in this area.
For this study, we take the HIJING simulation results shown
in Fig.\ref{fig:hijing_full} as an indication that
the single component neutral cluster model is
also a good {\em hadronic} model of
$AA$ collisions and explore the consequences.
A case study with a few hadronic event generators is under way and
will be reported elsewhere.

With only single species of
clusters, the joint probability to have an event with
particle rapidities $(\{y_a\})$ is simply
  \begin{eqnarray}
  \rho(\{y_a\})=
  \prod_{k=1}^{M_0}f_0(y_k^+, y_k^-) \,.
  \end{eqnarray}
All observables in this model depend only on the pair
distribution function $f_0(y^+,y^-)$. Hereafter, we will drop the
subscript
$0$ from $f_0$ for brevity.
This is pertinent to the
heavy-ion collisions at RHIC where the net charge is almost zero in the
central region.
For completeness, we have also listed the formulae for unpaired
net charges in the appendix.  As discussed in the appendix, the
contribution
of the non-zero net charges to the charge transfer fluctuations
is negligible as long as $Q \ll N_{\rm ch}$.

Due to our assumption
that the decay products are identical except for the electrical
charges, the 2-point function has the following symmetries
  \begin{subequations}
  \label{eq:charge:pairfunc-symmetry}
  \begin{eqnarray}
   f(y^+,y^-) &=&  f(y^-,y^+) \,  ,
   \label{eq:charge:pairfunc-symmetry1}
   \\
   f(y^+,y^-) &=&  f(-y^+,-y^-) \, .
   \label{eq:charge:pairfunc-symmetry2}
  \end{eqnarray}
 \end{subequations}
We will use these symmetry properties to simplify the equations for the
charge transfer fluctuations and the net charge fluctuations.

Guided by our discussion of the Thomas-Chao-Quigg relationship,
we use a separable form of pair
distribution function $f(y^+,y^-)=F(Y)R(y_{\rm rel})$, where $Y=(y^+ +
y^-)/2$
and $ y_{\rm rel}= y^+ - y^- $ as in Eqs.(\ref{eq:ChTr:f0-ansatz}) and
(\ref{eq:ChTr:f0-generalform}). The function $F(Y)$ is the rapidity
distribution of the clusters and
the function $R(y_{\rm rel})$ is the
relative rapidity distribution of the produced pair.
For the shape of $F(Y)$,
we use a Wood-Saxon form and a gaussian form here but more
sophisticated forms are certainly possible.
For $R(y_{\rm rel})$, the following two physically motivated
forms were used in our study:
  \begin{subequations}
  \label{eq:OneComp:Rr-two}
  \begin{eqnarray}
  R(r) &=&  \frac{1}{\gamma \sqrt{2\pi}} \,
    \exp{ \left({-r^2}/{2\gamma^2}\right) }
  \label{eq:OneComp:Rr-gaussian}
  \\
  R(r) &=& \frac{1}{2 \gamma} \,
    \exp{ \left({-|r|}/{\gamma}\right) }
  \label{eq:OneComp:Rr-exponential}
  \end{eqnarray}
  \end{subequations}
The function in Eq.(\ref{eq:OneComp:Rr-exponential}) is
the same as
Eq.(\ref{eq:ChTr:f0-generalform}) with $\gamma = 2
\kappa$.

The combination of $F(Y)$
and $R(r)$ are not arbitrary. Once we fix the charge correlation
length $\gamma$ for a particular $R(r)$, then the parameters for $F(Y)$
are
complete determined by the best fit to the experimental charged particle
distributions \cite{Back:2001xy,Wozniak:2004kp}.
To test the sensitivity to the different forms of the correlation
function,
we use the following 4 parameter sets in this section:
The parameter set labeled $1$ uses a Wood-Saxon form of $F(Y)$ and a
gaussian $R(y_{\rm rel})$.  The set labeled $2$ uses a Wood-Saxon $F(Y)$
but
$R(y_{\rm rel})$ has the exponential from. The parameter set labeled $3$
uses a gaussian $F(Y)$ and also a gaussian $R(y_{\rm rel})$.
The set labeled $4$ uses a gaussian $F(Y)$ but
$R(y_{\rm rel})$ has the exponential form.
These will be used for the net
charge fluctuation analysis and the charge transfer fluctuation analysis.

\subsection{Net Charge Fluctuation
\label{sect::one-comp:subsect:netchargefluc}}

STAR collaboration at RHIC has published their measurement of net charge
fluctuations \cite{Adams:2003st} and concluded that their result is
consistent with the hadronic gas expectations.
In this section, we use this data to fix the correlation length $\gamma$.
Since our previous fit to the HIJING simulations
gave $\gamma \sim 1.3 - 1.4$,
we will look for the correlation length within the range of
$1 < \gamma < 2$.

Within the observation window $-y_{\rm o} < y < y_{\rm o}$, the net
charge and the charge transfer are given by
  \begin{eqnarray}
  u(y) &=& \left[ Q_F(y)-Q_B(y) \right] /2
  \nonumber \\
  Q(y_{\rm o}) &=& \left[ Q_F(y) + Q_B(y)
  \right]
  \label{eq:OneComp:chargetrans-netcharge}
  \end{eqnarray}
where now $Q_F(y)$ and $Q_B(y)$ are measured within the observational
window.
Notice that $Q(y_{\rm o})$ is actually independent of $y$.
In the limit $y= y_{\rm o}$, we have $u= Q/2$.

In terms of the charge pair distribution function, the net charge
fluctuation is given by
  \begin{eqnarray}
  \delta Q^2(y_{\rm o})
  &=& \langle Q(y_{\rm o})^2 \rangle - \langle Q(y_{\rm o}) \rangle ^2
  \nonumber\\
  &=& 4 \langle M_0 \rangle
  \int_{-\infty}^{-y_{\rm o}}{dy^-  \int_{-y_{\rm o}}^{y_{\rm o}}
  {dy^+ f(y^+,y^-) } }
     \, .
  \label{eq:OneComp:NetCharge-useSymmetry}
  \end{eqnarray}
In deriving the above equation, we have made full use of the symmetry
properties
of the pair distribution function,
Eqs.(\ref{eq:charge:pairfunc-symmetry}).
Since the two particle distribution function $f(y^+,y^-)$ is peaked at
$|y^+ - y^-| = 0$, the net charge fluctuations in
Eq.(\ref{eq:OneComp:NetCharge-useSymmetry}) measure the local correlations
at
around the edge of the observation window $y \sim \pm y_{\rm o}$.
One can also see that as $y_{\rm o}\to \infty$, the net charge fluctuation
vanishes as it must.

The total number of charged particles
in the given rapidity window $(-y_{\rm o} , y_{\rm
o})$ is:
  \begin{eqnarray}
  {N_{\rm ch}(y_{\rm o})} = {2 \langle M_0 \rangle}
   \int_{-y_{\rm o}}^{y_{\rm o}}{dy^-  \int_{-\infty}^{\infty}
   {dy^+ f(y^+,y^-)} }   \, .
  \label{eq:OneComp:TotalChargeNumber}
  \end{eqnarray}
The ratio of the net charge fluctuations to the total number of charges in
the
rapidity window $(-0.5,0.5)$ was measured in the RHIC experiments, and the
value
for central collisions is around $\delta Q^2/N_{\rm ch}= 0.8-0.9$ after
correcting
for global charge conservation effect \cite{Adams:2003st,Adcox:2002mm}.
The correction is made through a
factor of $1/(1-p)$ where $p$ is the fraction of the charged particles
included in the observation compared to
the total number of charged particles produced.
>From this value, we can find
the correlation length for charged pairs.
The exact value depends slightly on the assumption on the shape of the
pair
correlation function $R(r)$ in relative rapidity space and the charge
center
distribution $F(Y)$.

\begin{figure}[t]
\includegraphics[scale=0.30,angle=90]{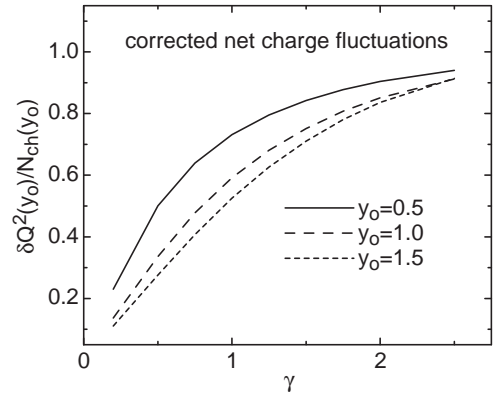}
\caption{
The charge conservation corrected ratio of the net charge fluctuations
to the total number of charged particles
$\delta Q^2(y_{\rm o})/N_{\rm ch}(y_{\rm o})$
is shown as a function of the charge correlation length $\gamma$.
We show the ratio at three different rapidity observation windows,
$|y|< y_{\rm o}$, where $y_{\rm o}=0.5$, $1.0$ and $1.5$ for the three
lines respectively. The parameters for the charge pair center distribution
function $F(y)$ are adjusted to fit the charged particle distribution data
measured by PHOBOS group at $\sqrt{s}=130$ GeV.
\label{fig:netChFluc}}
\end{figure}

\begin{figure}[t]
\center
\includegraphics[scale=0.30,angle=90]{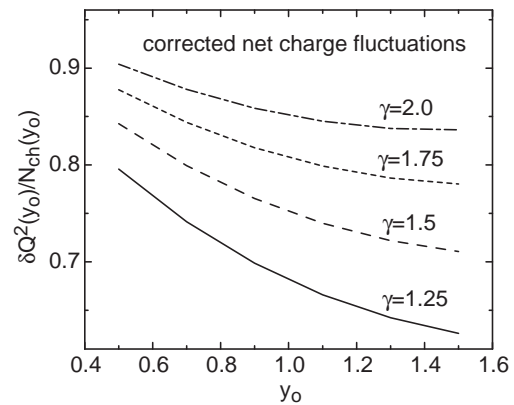}
\caption{
The corrected net charge fluctuation ratio
$\delta Q^2(y_{\rm o})/N_{\rm ch}(y_{\rm o})$
is shown as a function of the observation window size $y_{\rm o}$ for
different values of charge correlation lengths $\gamma=1.25$, $1.5$,
$1.75$ and $2.0$ respectively. The net charge fluctuations are decreasing
functions of observation window.
\label{fig:ChFluc-y} }
\end{figure}

\begin{figure}[t]
\center
\includegraphics[scale=0.30,angle=90]{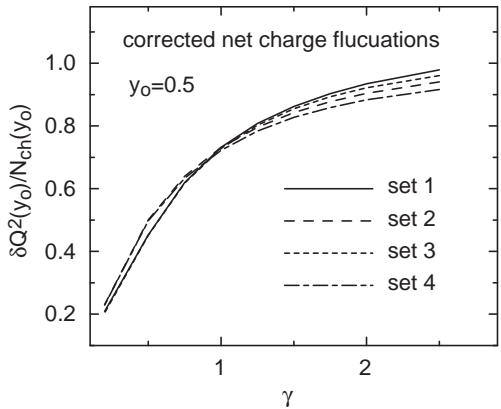}
\caption{
The charge conservation corrected ratio of the net charge fluctuations
to the total number of charged particles is plotted as a function of
the charge correlation length $\gamma$. Set $1$ refers to a Wood-Saxon
form for the charge pair center distribution $F(y)$, and a gaussian form
for the relative rapidity distribution between the pair $R(r)$. Set 2
refers to a Wood-Saxon form for $F(y)$ and an exponential decay form for
$R(r)$. Set 3 and 4 are for gaussian form of $F(y)$ with gaussian or
exponential decay form of $R(r)$ respectively. This should be compared
with data reported for  $\sqrt{s}=130$ GeV Au-Au collisions at RHIC.
Here the observation window is $(-0.5,0.5)$ in rapidity.
\label{fig:corrChFluc-mul} }
\end{figure}

In Fig.\ref{fig:netChFluc}, we have plotted the ratio of the net charge
fluctuations to the
total number of charges $\delta Q^2(y_{\rm o})/N_{\rm ch}(y_{\rm o})$
as a function of the pair correlation length $\gamma$.
We only show here
the charge fluctuation results from the parameter set $2$, where the
charge center distribution $F(Y)$ is Wood-Saxon form and the relative
rapidity distribution between the pair $R(r)$ is exponential decay form.
The different assumed
forms of the pair distribution function
$f = F(y)R(r)$ have little effect on the net charge fluctuations and are
not show here for clarity. From this figure, one can conclude
that the net charge fluctuations $\delta Q^2(y_{\rm o})$
are strongly
correlated with the charge correlation length $\gamma$ in this single
component case.

RHIC experiments can measure the net charge fluctuations as a
function of the observational window size $y_{\rm o}$. We have plotted
the corrected net charge fluctuations as a function $y_{\rm o}$
in Fig.\ref{fig:ChFluc-y}. As can be seen in this figure, the net charge
fluctuations always decrease when the observation window is enlarged.
This is because the total number of charged particles included in the
observation window is increasing faster than the net charge fluctuations.
The slope however is related to the charge correlation length.

Using the corrected net charge fluctuation ratio data from STAR
\cite{Adams:2003st}, we deduce that the charge correlation length is
  \begin{equation}
  \gamma \approx 1.5
  \end{equation}
This deduced value
is largely independent of the shape of the charge pair correlation
function $R(r)$ or the pair center distribution function $F(Y)$ as shown
in Fig.\ref{fig:corrChFluc-mul}. Notice that the inferred charge
correlation
length is consistent with the HIJING results where
$\kappa=\gamma/2 \approx 0.7$ in the previous simple estimate within the
rapidity window $(-1, 1)$.

\subsection{Charge Transfer Fluctuations
\label{sect::one-comp:subsect:chargetrans}}

We now apply our model to the charge transfer fluctuations
using the same parameter sets for the pair distribution function
$f(y^+,y^-)$ as we have used in the last section.

First, we need to express the the charge transfer fluctuations in terms of
the
pair distribution function $f(y^+,y^-)$.
Making full use of the symmetries
of the function $f(y^+,y^-)$ given
in Eq.(\ref{eq:charge:pairfunc-symmetry}), we
can simplify Eq.(\ref{eq:ChTr:Duy-general}) to
  \begin{eqnarray}
  \bar{D}_u(y) &=&
   { \delta Q^2(y_{\rm o}) \over 4}
   -\ave{\delta Q_F(y)\,\delta Q_B(y)}
   \non
   & = &
   { \delta Q^2(y_{\rm o}) \over 4}
   +2 \langle M_0 \rangle \int_{-y_{\rm o}}^{y}
   {dy^- \int_{y}^{y_{\rm o}}{dy^+  f(y^+,y^-)} }
   \nonumber\\
  \label{eq:OneComp:ChargeTrans-useSymmetry}
  \end{eqnarray}
where we used Eq.(\ref{eq:OneComp:NetCharge-useSymmetry}) and defined
$\delta X \equiv X - \ave{X}$.
Written this way, it is clear
that the charge transfer fluctuations depend on
both the charge correlations at
the edges of the observation window $y \sim \pm y_{\rm o}$ and at the
forward-backward rapidity cut $y$.
From this expression, one also sees that
the lower limit of charge transfer fluctuations
is $\{\bar{D}_u(y)\}_{\rm min}=\delta Q^2(y_{\rm o})/4 $.

\begin{figure}[t]
\center
\includegraphics[scale=0.30,angle=90]{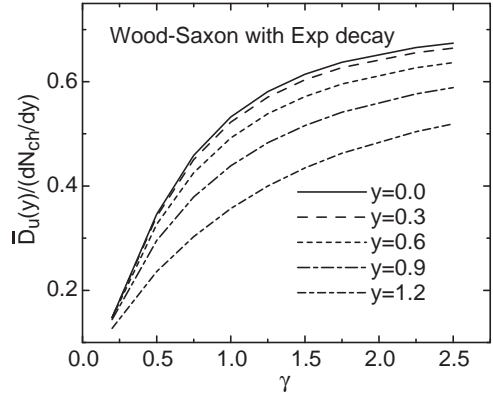}
\caption{
The ratio of the charge transfer fluctuations $\bar{D}_u(y)$ to the
charged particle yield $dN_{\rm ch}/dy$ in the observation window
$(-1.5,1.5)$ is plotted as a function of the charged particle correlation
length $\gamma$. The different lines represent the results for different
forward-backward rapidity cut, $y=0.0$, $0.3$, $0.6$, $0.9$ and $1.2$
respectively. We only showed the results from parameter set $2$, while
other parameter sets yield qualitatively similar results. This result is
for the one component model.
\label{fig:ChTransfer} }
\end{figure}

In Fig.\ref{fig:ChTransfer}, we have plotted the ratio of the charge
transfer fluctuation $\bar{D}_u(y)$ to the charged particle yield
$dN_{\rm ch}/dy$ as a function of pair correlation length $\gamma$.
As the observed rapidity spectrum is nearly flat around midrapidity,
dividing by $N_{\rm ch}$ or $dN_{\rm ch}/dy$ only affects the overall
scale.
The value of the ratio strongly
depends on the correlation length $\gamma$.
Also the ratio decreases as the forward-backward
separation $y$ increases at a fixed $\gamma$.
As mentioned before (c.f.~Eq.(\ref{eq:ChTr:Fy-flat-solveDuy})),
this decrease  is due to the limited observation window and
has nothing to do with the changing correlation length.
This can be also shown in the following way.
If $dN_{\rm ch}/dy$ does not vary significantly within the
observational window, then $F(Y)$ does not vary significantly
within the observational window. In that case,
\begin{eqnarray}
{1\over \langle{M_0}\rangle} {\partial \bar{D}(y)\over \partial y}
\approx
-F(0) \int_{y_{\rm o}-y}^{y_{\rm o}+y} dr\, R(r) < 0
\end{eqnarray}
for $y> 0$.
Therefore, if $dN_{\rm ch}/dy$
is flat {\em and} the system is composed
of only one species of neutral clusters, $\bar{D}_u(y)$ must be a
decreasing function of $y > 0$. Conversely, if
$\bar{D}_u(y)$ is an {\em increasing} function of $y> 0$
while $dN_{\rm ch}/dy$ remains flat, it signals
the existence of a new component.  We turn to this possibility in the
next section.

\section{Two Component Model \label{sect::two-comp}}

In a two component model, the full joint probability for the
rapidities is given by
  \begin{eqnarray}
  \rho(\{y_i\})
  =
    \prod_{j=1}^{M_1}f_{1}(y_j^+, y_j^-) \, 
    \prod_{k=1}^{M_2}f_{2}(y_k^+, y_k^-) \,, 
    \label{eq:two_comp_rho}
  \end{eqnarray}
where $f_{1}$ and $f_{2}$ have different correlations lengths
$\gamma_{1} > \gamma_{2}$. The pair distribution functions are again
taken as the separable form: $f_i(y^+,y^-)=F_i(Y)R_i(r)$
where $Y=(y^+ +y^-)/2$ and $r=y^+ - y^-$.
Since the fluctuations add in quadrature, the net charge fluctuations
and the charge transfer fluctuations are just the sum of contributions
from the two components, $f_{1}$ and $f_{2}$ respectively.

Physically, the two component model is motivated by the fact that the hot
and
dense matter produced in the relativistic heavy-ion reaction is not
necessarily
homogeneous in rapidity space as explained before.
If the deconfined QGP phase did exist during the
early stage of heavy-ion reaction, it is highly possible that the QGP
phase
coexisted with the hadron gas phase. A simple situation would be a phase
separation between the QGP phase and the hadron gas phase. This could
produce
signals that are specific to a phase coexistence scenario.
In the case of the charge transfer fluctuations, a QGP and hadron gas
phase
separation could be measured and mapped into the charge correlations in
the
relative rapidity space between the pair of particles.
In the following, we will refer to the short correlation
part as a `QGP' and the long correlation part as a `hadron gas' (HG).

\begin{figure}[t]
\center
\includegraphics[scale=0.3,angle=90]{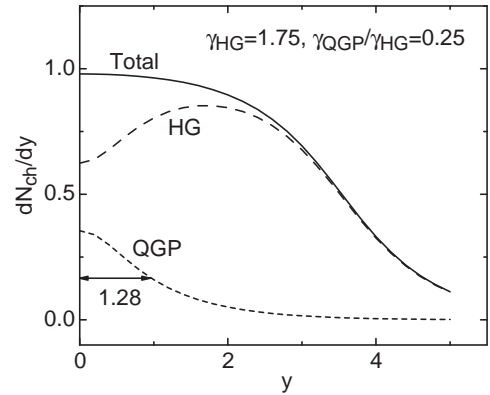}
\caption{
The Charge yields are plotted as a function of rapidity in a two component
system. The full line represent the total charged particle yield profile
while
the dashed and dotted lines represent the contributions from the hadron
gas
and the QGP phases in a typical two component model we used. The QGP size
is
$\xi=1.28$ in this case.
\label{fig:TwoComp:ChargeProfile} }
\end{figure}

In our simple two component model, we assume the two components have
different correlation lengths and the $R(r)$ functions
are either taken as an exponential form or a gaussian form,
with the corresponding correlation lengths satisfying
$\gamma_1 > \gamma_2$.
Here $\gamma_1 = \gamma_{\rm HG}$ is the rapidity correlation length of
the
hadronic part (labeled `1') and $\gamma_2 = \gamma_{\rm QGP}$ is that of
the
QGP part (labeled `2').
We let the cluster distribution functions
for the two components be:
  \begin{subequations}
  \label{eq:TwoComp:pairfunc-twocomp}
  \begin{eqnarray}
   \langle{M_1}\rangle
   F_1(y) &=&  \frac{c_1}{1+\exp{\left[{(|y|-\sigma_0)}/{a_0}\right]}}
            - c_2  g_{1}^{\vphantom{x}}(y) \,,
   \nonumber \\
   \label{eq:TwoComp:pairfunc-twocomp1}
   \\
   \langle{M_2}\rangle F_2(y) &=& c_2  g_2(y)\,.
   \label{eq:TwoComp:pairfunc-twocomp2}
  \end{eqnarray}
  \end{subequations}
Where $c_1$ is a normalization factor and $c_2$ is the strength of the
QGP phase.
To be physically consistent, the value of $c_2$ is adjusted
so that the function $F_1$ is always positive.
We then demand that $\int R_1 g_{1} = \int R_2 g_2$ so that
$dN_{\rm ch}/dy$ is independent of the choices of $g_1$
and $g_2$.
The parameters $\sigma_0$ and $a_0$ are chosen to fit the
experimental data.

For a gaussian $R_i(r)$,
it is convenient to take the functional forms of $g_1(y)$ and $g_2(y)$
to be also gaussian.
To satisfy $\int R_1 g_1 = \int R_2 g_2$,
the widths of these gaussians should be related as follows
$\xi^2=\gamma_1^2/4+\sigma_1^2=\gamma_2^2/4+\sigma_2^2$
where $\sigma_i$ is the width of $g_i(y)$ and $\xi$ is the width of the
QGP
part of $d\Nch/dy$.
For an exponential $R_i(r)$,
determining the forms of $g_1$ and $g_2$
is a little more complicated than the gaussian case.
However, since this form of the correlation function satisfies the
Thomas-Chao-Quigg relationship, we will use mainly the exponential form
hereafter. As in the one component model,
both the net charge fluctuation and the charge transfer fluctuation
results
are not very sensitive to the particular choice of the functional form
the two particle distribution function.

With the exponential
$R_i(r)$, the charge center distribution functions $g_1(y)$ and $g_2(y)$
can
not be both gaussian anymore. It is convenient to
select $g_1(y)$ (the hadronic part) to have  a gaussian form.
Using the fact that the function $R_i(r)$ is in fact a Green function
of the differential operator $(d/dr)^2 - 1/\gamma_i^2$,
the function $g_2(y)$ (the QGP part) can be obtained as
   \begin{eqnarray}
    g_2(y) = \left(\frac{\gamma_2}{\gamma_1}\right)^2 g_1(y)
   + \left[ 1-\left(\frac{\gamma_2}{\gamma_1}\right)^2 \right] \rho(y)
   \label{eq:TwoComp:exp-integral-transformation}
   \end{eqnarray}
where $\rho(y) = \int R_1 g_1 = \int R_2 g_2$ is
     \begin{eqnarray}
    \rho(y) =
    \int_{-\infty}^{\infty} dx \, g_1\left( \frac{x+y}{2} \right)
    \frac{1}{2\gamma_1} \exp \left(-\frac{|x-y|}{\gamma_1}\right)
     \label{eq:TwoComp:integral-to-rho}
   \end{eqnarray}
The function $\rho(y)$ is
proportional to $dN_{\rm ch}/dy$ of the QGP part.
The width $\xi$ of the QGP part is determined by the width of $\rho(y)$.

In all, we have 8 parameters here.
We have $c_1$, $\sigma_0$, $a_0$ and $c_2$
explicitly appearing in Eq.(\ref{eq:TwoComp:pairfunc-twocomp}).
The parameter $\sigma_1$ is the width of the function $g_1$.
The parameter $\xi$ is the width of the function
$\rho(y)$ and is connected with the parameter $\sigma_1$ through
Eq.(\ref{eq:TwoComp:integral-to-rho}).
We also have two charge correlation lengths $\gamma_1$ and $\gamma_2$
with the condition $\gamma_1 > \gamma_2$.

Among the 8 parameters, 5 are fixed in the following way.
Since we are only interested in the ratio $D_u(y)/(dN/dy)$, the value of
parameter $c_1$ is irrelevant and we just fix it to be 1.
The parameters $a_0$ and $\sigma_0$ are fixed by requiring that the
resulting $dN/dy$ shape describes results from RHIC experiments.
The parameter $c_2$ is always chosen to be the maximum possible value
for the condition $F_1(y)\ge 0$ given all other parameters.
The parameter $\sigma_1$ is determined by the parameter $\xi$.

The three parameters we are going to vary in the following
are then $\gamma_1$, $\gamma_2$ and $\xi$.
An example of total charged particle spectrum with
the respective hadron gas and QGP contributions
are plotted in Fig.\ref{fig:TwoComp:ChargeProfile}.
There is a substantial presence of QGP
around midrapidity, but it becomes less prominent for $|y| > \xi$.

\subsection{Net Charge Fluctuation
\label{sect::two-comp:subsect:chargefluc}}

 As in the one component model, we first consider the net charge
 fluctuations to further fix our parameters.
\begin{figure}[t]
\center
\includegraphics[scale=0.3,angle=90]{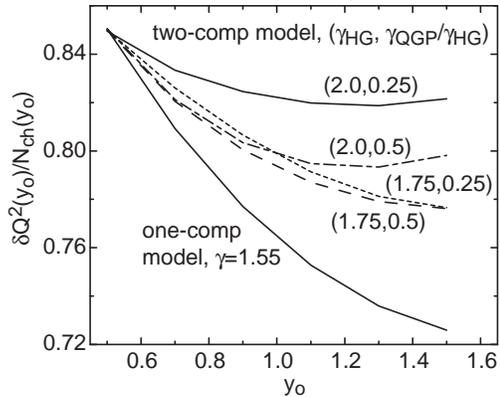}
\caption{
The ratios of the corrected net Charge fluctuations to the total number
of charges $\delta Q^2(y_{\rm o})/N_{\rm ch}(y_{\rm o})$ in the rapidity
observation window $(-y_{\rm o}, y_{\rm o})$ are plotted as a function
of rapidity $y_{\rm o}$ for the one and two component models. The line
for the one component model is labelled by the charge correlation length
$\gamma=1.44$, and the lines for the two component models are labelled
by the pair of charge correlation lengths in the two components
$(\gamma_{\rm HG},\gamma_{\rm QGP}/\gamma_{\rm HG})$.
The corrected net charge fluctuations are fixed
$\delta Q^2(0.5)/N_{\rm ch}(0.5)=0.85$ for all lines.
 \label{fig:TwoComp:Chargefluc} }
\end{figure}
When there are two distinct species of neutral clusters, the net charge
fluctuation within the rapidity interval $(-y_{\rm o}, y_{\rm o})$ is
given by
  \begin{eqnarray}
  \lefteqn{\delta Q^2(y_{\rm o})
  =
  4 \langle M_1 \rangle
  \int_{-\infty}^{-y_{\rm o}}{dy^-  \int_{-y_{\rm o}}^{y_{\rm o}}
  {dy^+ f_1(y^+,y^-) } }} &&
  \non
  & & {}
  +
  4 \langle M_2 \rangle
  \int_{-\infty}^{-y_{\rm o}}{dy^-  \int_{-y_{\rm o}}^{y_{\rm o}}
  {dy^+ f_2(y^+,y^-) } }
  \label{eq:TwoComp:NetCharge-useSymmetry}
  \end{eqnarray}
and the total charged multiplicity is given by
\be
\lefteqn{N_{\rm ch}(y_{\rm o})
= 2\ave{M_1}\int_{-y_{\rm o}}^{y_{\rm o}} dy^-
\int_{-\infty}^\infty dy^+\, f_1(y^-, y^+)
} &&
\non
& & {}
+
2\ave{M_2}\int_{-y_{\rm o}}^{y_{\rm o}} dy^-
\int_{-\infty}^\infty dy^+\, f_2(y^-, y^+)
\ee

The ratios
$\delta Q^2(y_{\rm o})/N_{\rm ch}(y_{\rm o})$ corrected as before by a
factor of $1/(1-p)$
are plotted as a function of the observation box size $y_{\rm o}$ in
Fig.\ref{fig:TwoComp:Chargefluc}. The lowest solid line is for the one
component model with charge correlation length $\gamma=1.55$ as obtained
in
the last section.
For the two component model,
4 different choices with $\gamma_{1}=\gamma_{\rm HG} = 1.75, 2.0$ and
$\gamma_{2}/\gamma_{1}=\gamma_{\rm QGP}/\gamma_{\rm HG} = 0.25, 0.5$ are
shown.
Since the corrected net charge fluctuations
at $y_{\rm o}=0.5$ is about $0.8-0.9$ \cite{Adams:2003st,Adcox:2002mm},
the QGP width parameter $\xi$ are chosen in such a way that
all net charge fluctuations have
$\delta Q^2(y_{\rm o})/N_{\rm ch}(y_{\rm o}) = 0.85$ at $y_{\rm o}=0.5$.
For instanec, for the parameter set
$\gamma_{\rm HG} = 1.75$ and $\gamma_{\rm QGP}=
\gamma_{\rm HG}/4 = 0.44$, the width of the QGP part $\xi$ turned out to
be
$\xi = 1.28$.

The net charge fluctuations as a function of rapidity is flatter in
the two component model than in the one component model. This is because
the two component results interpolate between the one component results
with $\gamma=\gamma_{\rm QGP}$ and $\gamma=\gamma_{\rm HG}$.
Since the behaviors of the net charge fluctuations
for the one component model and the two component model are clearly
distinct, one is tempted to argue that the
flat $\delta Q^2(y_{\rm o})/N_{\rm ch}(y_{\rm o})$ itself is an indication
of a second phase. (A similar idea was suggested in
Ref.\cite{Fialkowski:2001zq}.)
Unfortunately, totally uncorrelated system also has a
flat $\delta Q^2(y_{\rm o})/N_{\rm ch}(y_{\rm o})$ when corrected for the
effect of total charge conservation.

In addition,
the net charge fluctuation $\delta Q^2(y_{\rm o})/N_{\rm ch}(y_{\rm o})$
is constrained by the fact that in the limit $y_{\rm o}\to 0$,
we should get the Poisson limit
$\delta Q^2(y_{\rm o})/N_{\rm ch}(y_{\rm o}) \to 1$.
This puts a constrain on the sensitivity of net charge
fluctuations to the QGP phase.
In our two component model, the QGP phase
is located mostly around midrapidity, and the presence of QGP is
reduced at larger rapidities. Hence in order
to observe the QGP, we need to have $y_{\rm o} \sim 0$.
But because of the limiting value
at $y_{\rm o}=0$, the net charge fluctuations actually have a reduced
sensitivity to the QGP phase.
The charge transfer fluctuation, which we now turn our attention to, does
not
have these limitations.

\subsection{Charge Transfer Fluctuation
\label{sect::two-comp:subsect:chargetrans}}

The charge transfer fluctuation $\bar{D}_u(y)$ is qualitatively different
than the net charge fluctuation $\delta Q^2(y_{\rm o})$.
As will be shown in this section, the charge transfer
fluctuations is capable of distinguishing
the two phases of our two component model and
hence can be used as a signal for the QGP phase.
In our model, the charge transfer fluctuation $\bar{D}_u(y)$ is given
by
  \begin{eqnarray}
  \bar{D}_u(y) &=&
  {\delta Q^2(y_0)\over 4}
   +2 \langle M_1 \rangle \int_{-y_{\rm o}}^{y}
   {dy^- \int_{y}^{y_{\rm o}}{dy^+  f_1(y^+,y^-)} }
   \nonumber\\
   && +2 \langle M_2 \rangle \int_{-y_{\rm o}}^{y}
   {dy^- \int_{y}^{y_{\rm o}}{dy^+  f_2(y^+,y^-)} }
  \label{eq:twoComp:ChargeTrans-useSymmetry}
  \end{eqnarray}
while the final particle spectrum is
\be
{dN_{\rm ch}\over dy}
=
   2\langle M_1 \rangle h_1(y)
   +
   2\langle M_2 \rangle h_2(y) \, .
\ee
where $h_i(y) = \int_{-\infty}^\infty dx\, f_i(x, y)$.

For small $\gamma_i$, it is easy to show that
\begin{eqnarray}
  \bar{D}_u(y) \sim \hbox{const} +
   \gamma_1 \langle M_1 \rangle F_1(y)
   +
   \gamma_2 \langle M_2 \rangle F_2(y)
  \end{eqnarray}
while the rapidity spectrum becomes
\be
{dN_{\rm ch}\over dy} \sim
\langle M_1 \rangle F_1(y)
+
\langle M_2 \rangle F_2(y) \, .
\ee
Hence, changes in the charge transfer fluctuation compared to the charged
particle spectrum
reflect changes in the concentration of the two components and/or
the change in the mean correlation length.
When the correlation lengths are not very small,
$\bar{D}_u(y)$ is given in terms of the convolution of
$F_1(y)$ and $F_2(y)$ with
the corresponding relative distribution $R_1(r)$ and $R_2(r)$.
Unless $\gamma$'s are very large, the ratio
$\bar{D}_u(y)/(d\Nch/dy)$ should still be
sensitive to the changes in the composition.

If the net charge fluctuation $\delta Q^2(y_{\rm o})$ is sizable, its
presence can reduce the sensitivity of the ratio
\be
\bar{\kappa}(y) = {\bar{D}_u(y)\over (d\Nch/dy)}
\ee
to the changing composition since $d\Nch/dy$ abruptly decreases beyond the
central plateau. However, since this
$\delta Q^2(y_{\rm o})$ is the {\em uncorrected} net charge fluctuation,
it
is easy to measure and subtract it from $\bar{D}_u$.
In this case, the relevant ratio becomes
\be
\tilde{\kappa}(y) =
{\bar{D}_u(y) - \delta Q^2(y_{\rm o})/4\over (d\Nch/dy)}
\label{eq:corrected_Du2}
\ee
If $d\Nch/dy$ is flat within the observational window, this is of course
not
necessary as $\delta Q^2(y_{\rm o})$ term just adds a constant.
Also in the large $y_{\rm o}$ limit,
$\delta Q^2(y_{\rm o})\to 0$ due to
the overall charge conservation and hence this modification is not
necessary.

\begin{figure}[t]
\center
\includegraphics[scale=0.3,angle=90]{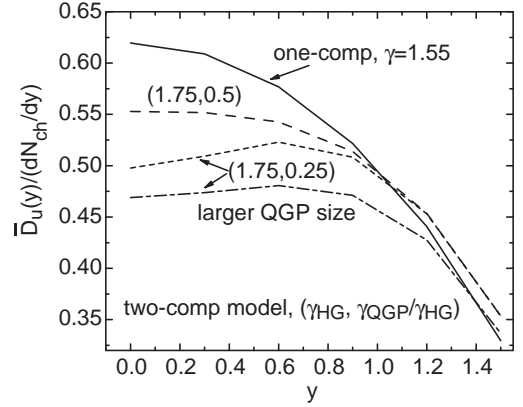}
\caption{
The ratio of the Charge transfer fluctuations to the total number
of charges $\bar{D}_u(y)/(dN_{\rm ch}/dy)$ in the rapidity observation
window $(-1.5, 1.5)$ is plotted as a function of the forward-backward
separation cut $y$ in the one component and two component models.
For
the one component model, the charge correlation length is fixed at
$\gamma=1.55$. For the two component models, the pair of correlation
lengths $(\gamma_{\rm HG}, \gamma_{\rm QGP}/\gamma_{\rm HG})$ are labelled
on each line. For the middle two lines, the QGP sizes $\xi$ are fixed by
requiring the net charge fluctuations to be constant $\delta
Q^2(0.5)/N_{\rm ch}(0.5)=0.85$. The two lines with the same charge
correlation lengths
$(\gamma_{\rm HG}, \gamma_{\rm QGP}/\gamma_{\rm HG})=(1.75,0.25)$
have different QGP size.
 \label{fig:TwoComp:ChargeTrans} }
\end{figure}

The quantity $\bar{\kappa}(y)$ within $-1.5 < y < 1.5$
are plotted in Fig.\ref{fig:TwoComp:ChargeTrans} as a function
of the forward-backward rapidity separation $y$ in the one and
two component models.
As our $d\Nch/dy$ is almost flat within this window, we don't need to
subtract $\delta Q^2(y_{\rm o})$ part.
The shape for the one component model is completely
fixed by the charge correlation length $\gamma=1.55$ as before,
and it is a decreasing function of $y$.
For the dashed and the dotted lines, we use the same parameters
as obtained in the last section based on the experimentally observed net
charge fluctuations.
Even though the the one and two component cases
have a common net charge fluctuation at $y_{\rm o}=0.5$, the charge
transfer fluctuation patterns are quite different:
The most prominent feature for the two component model is the appearance
of
the minimum for $\bar{\kappa}(y)$ at $y = 0$ for
$\gamma_{QGP} < 0.5\gamma_{HG}$,
while the single component case always has a maximum at $y = 0$.

The minimum appears at midrapidity because that is where the QGP component
is concentrated. As $y$ increases, the fraction of the QGP matter
decreases.  Hence, $\bar{\kappa}(y)$
increases as a function of $y > 0$.
The point where the slope of $\bar{\kappa}(y)$
changes sign must is directly related to the width of the QGP
component.  Unfortunately, when the size of the observation window and the
width of the QGP component are similar, the sensitivity
to the size of QGP is partially lost because $\bar{\kappa}(y)$ must
decrease as $y$ approaches the edge.
To measure the size of the QGP component well,
one needs to have $y_{\rm o} \gg \xi$.
This prompts us to extend the rapidity window of our
observation in the two component calculations.

\begin{figure}[t]
\center
\includegraphics[scale=0.3,angle=90]{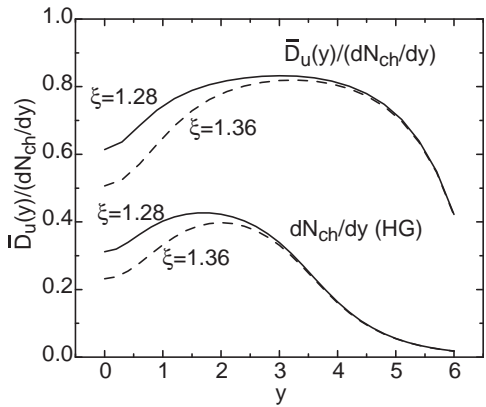}
\caption{
The charge transfer fluctuations $\bar{D}_u(y)/(dN_{\rm ch}/dy)$ are
shown as a function of the forward-backward rapidity cut $y$ in the
two component system.
The QGP size is indicated by
the parameter $\xi$. We have also plotted the corresponding hadron gas
charge profiles, $dN_{\rm ch}/dy(HG)$, for the two sets of
calculations.
  \label{fig:TwoComp:ChargeTrans-cut} }
\end{figure}

In Fig.\ref{fig:TwoComp:ChargeTrans-cut},
we plot
$\bar{\kappa}(y)$ with two different $\xi$'s as a function of
$y$.  The observation window used is
$(-6.0,6.0)$ which are large compared to the size of the QGP component.
One can conclude that the width of the depression around midrapidity
does reflect the width of the QGP component.
Nearly flat $\bar{\kappa}(y)$
for $y > |\xi|$ is expected since in this region
we should recover the single component result.  The decrease near the edge
is again due to the finite window size.
Also shown are the shapes of
the HG contribution to the rapidity spectrum as a reference.
The charge transfer fluctuations with the same two $\xi$'s as above
are also shown in
Fig.\ref{fig:TwoComp:ChargeTrans}. In the case of a more limited window,
the sensitivity to the QGP size is reduced due to the
fact that the size of $y_{\rm o}$ is in fact about the same as the width
of
the QGP part $\xi$.

The charge transfer fluctuation
$\bar{D}_u(y)$
is completely different from the net charge fluctuation
$\delta Q^2(y_{\rm o})$ as far as
the observation window size effect is concerned.
As discussed before,
having a larger observation window cannot increase the
sensitivity of the net charge fluctuations to the QGP phase when it is
confined to a small region around midrapidity.
However, for charge transfer fluctuations,
having a large observation window increases the sensitivity
since the window now encompasses more of the QGP part.
Enlarging the observation window also reduces the edge effect increasing
the
sensitivity even more.
An additional advantage for the charge transfer fluctuations
is that there is no the global charge conservation correction
unlike the net charge transfer fluctuations.

\begin{figure}[t]
\center
\includegraphics[scale=0.4]{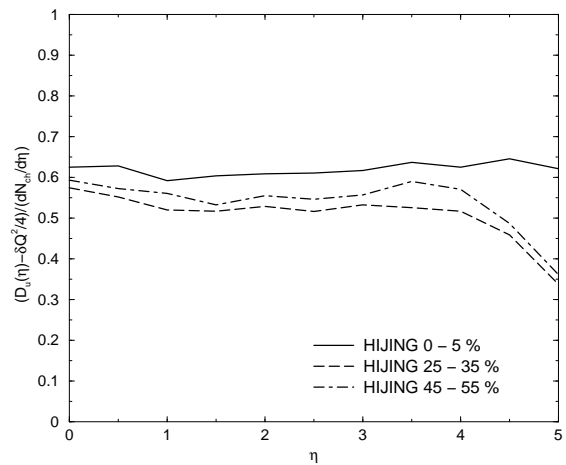}
\caption{
A plot of ratios $(D_u(\eta)-\delta Q^2/4)/(d\Nch/d\eta)$
for 3 different centrality
classes using 50,000 minimum bias HIJING events for
Au-Au collisions at
$\sqrt{s}=200\,\GeV$.
  \label{fig:hijing_ratio}}
\end{figure}

For a further reference, we show the result of analyzing 50,000 minimum
bias
HIJING events for Au-Au collisions at $\sqrt{s}=200\,\GeV$
in Fig.\ref{fig:hijing_ratio}
for 3 different centrality classes.
Again the analysis is carried out in the pseudo-rapidity space.
The data points used are the same as in Fig.\ref{fig:hijing_full}
except that the charge transfer fluctuations are corrected for the
overall net charge fluctuations as in Eq.(\ref{eq:corrected_Du2}).
The fact that $\delta Q^2 \ne 0$ even in the full phase space is
due to the spectators.  Since the net charge carried by
spectator nucleons fluctuates, so does the net charge of the produced
particles.  In the usual way of characterizing the centrality classes (by
$\Nch$ or $E_T$), this is unavoidable.
It should be observed that for all centrality classes, the ratio
$\tilde{\kappa}(\eta)=(D_u(\eta)-\delta Q^2/4)/(dN_{\rm ch}/d\eta)$
is essentially
flat. It is somewhat surprising the Thomas-Chao-Quigg relationship
works well for HIJING events given the fact that
no `cluster' appears explicitly within HIJING.

\subsection{Forward-Backward Multiplicity Correlation}

In all the above considerations, the key points are:
(i) there are primordial clusters that produce multiple particles,
(ii) a local cut separates the phase space into two regions,
(iii) one can define observables that only count the
primordial clusters that have their decay products separated by the
local cut (c.f. Fig.\ref{fig:concept0}).
These points imply that such observables are sensitive
to the local properties around the cut.
Hence, if the nature of the `clusters' changes in different regions of
the phase space, then these observables can detect the changes.

In some experiments, such as PHOBOS at RHIC,
charge states of the produced particles cannot be determined.
In this case, charge transfer fluctuation cannot be used.  However, since
the essence of the current method is to have a cut that separates produced
particles, just measuring the forward-backward multiplicity
correlation
  \begin{eqnarray}
  w(\eta)&  = &
  \ave{N_F(\eta)\, N_B(\eta)}
  -
  \ave{N_F(\eta)}\ave{N_B(\eta)},
  \end{eqnarray}
may be enough to detect the change in the correlation length.
We explore this idea in the following.
Here $N_F(\eta)$ is the charged multiplicity in the region forward of
the pseudo-rapidity $\eta$ and $N_B(\eta)$ is the charged multiplicity in
the region backward of $\eta$.

In this section, we switch to the pseudo-rapidity $\eta$ since without
particle identification one cannot determine the rapidity $y$.  However,
in the current formulation of the problem, the only change this switch
introduces is that instead of rapidity correlation function
$f_i(y, y')$ we have the pseudo-rapidity correlation function
$f_i(\eta, \eta')$.

Using the two component model Eq.(\ref{eq:two_comp_rho}),
it is easy to show that
  \be
  w(\eta) &=&
  4\left( \ave{\delta M_1^2} - \ave{M_1} \right)
  \int_{\eta}^{\eta_{\rm o}} d\eta\,
  h_1(\eta)\, \int_{-{\eta_{\rm o}}}^\eta d\eta\, h_1(\eta)
  \non &&
  +
  4\left( \ave{\delta M_2^2} - \ave{M_2} \right)
  \int_{\eta}^{\eta_{\rm o}} d\eta\, h_2(\eta)\,
  \int_{-\eta_{\rm o}}^\eta d\eta\, h_2(\eta)
  \non
  & &
  + 2\ave{M_1} \int_{\eta}^{\eta_{\rm o}} d\eta'
  \int_{-\eta_{\rm o}}^\eta d\eta''\, f_1(\eta', \eta'')
  \non
  & &
  + 2\ave{M_2} \int_{\eta}^{\eta_{\rm o}} d\eta'
  \int_{-\eta_{\rm o}}^\eta d\eta''\, f_2(\eta', \eta'')
  \label{eq:two_comp_multi_trans_integralform}
  \ee
where the correlation functions $f_1$ and $f_2$ are functions of
pseudo-rapidities and we defined
\be
h_i(\eta) = \int_{-\infty}^\infty d\eta'\, f_i(\eta', \eta)
\ee
The non-trivial part of this expression is essentially the same as the
charge transfer fluctuations.  The sensitivity of this observable to the
changes in the pseudo-rapidity correlation length depends crucially
on the size of the first two terms containing $\ave{\delta
M_i^2}-\ave{M_i}$.
If the number fluctuations of
the clusters obey Poisson statistics, then these two terms vanish.
In that case, $w(\eta)$
is as sensitive as the charge transfer fluctuation to the
presence of the second phase.
In the limit where there is only a single species of clusters
and also $\eta_{\rm o} \to \infty$,
we have an additional Thomas-Chao-Quigg relationship
  \be
  w(\eta) \approx \kappa {dN_{\rm ch}\over d\eta}
  \ee
with a constant $\kappa$, provided that $\ave{\delta M^2} = \ave{M}$.

\begin{figure}[t]
\center
\includegraphics[scale=0.3,angle=90]{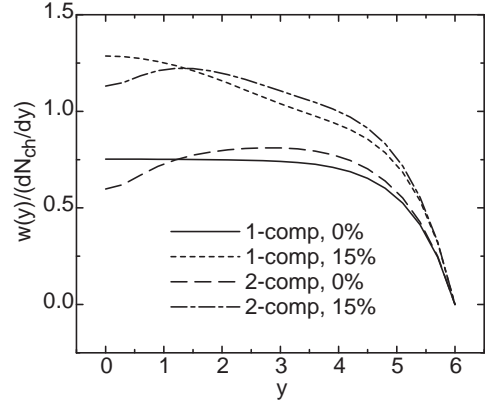}
\caption{
The ratio of the forward-backward multiplicity correlation $w(\eta)$ to
the charged particle yield $dN_{\rm ch}/d\eta$ is plotted is plotted
as a function of $\eta$ for one component and two component models.
The fluctuations of the total number of charged particles are also
indicated for the corresponding lines. For Poisson distribution of
$M_i$ ($i=1$ or $2$), the deviation from Poisson is $0\%$, that is,
$\langle \delta M_i^2 \rangle - \langle M_i \rangle = 0$. We have also
included the results for $15\%$ deviations.
  \label{fig:TwoComp:multi-transfer} }
\end{figure}

The ratio of the forward-backward multiplicity correlation $w(\eta)$ to
the
charged particle yield $dN_{\rm ch}/d\eta$ is plotted in
Fig.\ref{fig:TwoComp:multi-transfer}. When the clusters are
distributed according to Poisson distributions,
the Thomas-Chao-Quigg relationship
holds for a single component model and the ratio of
$w(\eta)/(dN_{\rm ch}/d\eta)$ is flat in the central region
(The solid line in Fig.\ref{fig:TwoComp:multi-transfer}).
For a two component model with Poisson statistics (long dashed line),
we see a minimum at midrapidity just as in the charge transfer
fluctuation.

When the statistics deviates from Poisson,
the factor
$\langle \delta M_i^2 \rangle - \langle M_i \rangle$ is non-zero.
Then the first two terms in
Eq.(\ref{eq:two_comp_multi_trans_integralform})
contributes.  These terms decrease with $\eta$ and hence partially
compensates the rising part due to the correlation change.
However,
the qualitative trend of the forward-backward multiplicity correlation
still remains valid: The increasing segment is still present
in the two component model and is an indication of the presence of QGP.

For heavy particles originating from a thermally equilibrated system, the
multiplicity fluctuation should follow the
Poisson statistics. For light particles,
$\ave{\delta M^2}$ can deviate up to $15\,\%$ from the Poisson value.
As an estimate, in Fig.~\ref{fig:TwoComp:multi-transfer} we show our
results
with $\langle \delta M_i^2 \rangle - \langle M_i \rangle =
0.15\,\ave{M_i}$.
One can still see a clear dip near midrapidity.

Coincidentally, the PHOBOS group has also measured a variation of
the signal we proposed here (see \cite{Wozniak:2004kp}).
The difference between the forward-backward multiplicity correlation
$w(\eta)$
and the ``charged particle multiplicity fluctuations"
$\sigma (C)$ used by
the PHOBOS group is subtle but results in quite different sensitivity.
The signal $\sigma (C)$ in \cite{Wozniak:2004kp} measures the correlations
between rapidity regions of
$(-\eta-\Delta\eta/2,-\eta+\Delta\eta/2)$
and $(\eta-\Delta\eta/2,\eta+\Delta\eta/2)$,
with each region covering the same
rapidity window of $\Delta\eta=0.5$, $1.0$ and $1.5$.
Typically these two rapidity
regions do not have a common edge.
In our case, the two pseudo-rapidity regions are
$(-\eta_{\rm o}, \eta)$ and $(\eta,\eta_{\rm o})$.
They share a common edge at $\eta$, but
the two regions are generally of different size.
When $\eta=\Delta\eta/2$, $\sigma (C)$ is the same as
the forward-backward multiplicity correlation $w(0)$.
However,
as we have shown in this paper, the single point in the fluctuation
measurement can not distinguish between one component and two component
models as the charge correlation length can be adjusted to fit this single
point.
One must measure $w(\eta)$ as a function of $\eta$ to get full information
on possible phase change.

\section{Summary \label{sect::summary}}

In this paper, we proposed the charge transfer fluctuations as a signal
for the QGP phase of matter.
The essence of our argument is very simple.
Suppose there are strong unlike-sign correlations in the underlying
system, then the charges are locally conserved.
With a separating wall in the local region where we want to explore,
the correlations (or fluctuations) of charges across this wall
are sensitive to the  {\em local} charge correlation length.
The pairs created far away from the
separating wall contribute little to the fluctuations since they have
little chance to be separated by the wall. This is essentially the idea
behind the charge transfer fluctuations proposed for earlier $pp$
collisions.

Since the charge transfer fluctuation $D_u(y)$ is a local measure of the
unlike-sign correlation length,
it can be used to detect the presence of a second phase in $AA$
collisions.
Quite generally, one can say that
the presence of a local {\em minimum} at $y=0$
for the ratio $\kappa(y) = D_u(y)/(d\Nch/dy)$
is a signal for the second (presumably QGP) phase.
The appearance of this minimum is due to the facts that
(i) a QGP phase should appear around midrapidity where the density is
the highest (ii) hadrons coming out of
a QGP phase should have a markedly short unlike-sign
correlation length compared to that of the hadronic matter.
The size of the depression near midrapidity in turn
contains information on the size of the QGP phase.  If one has a large
observation window, the extent of the second phase in the rapidity space
can
be in fact estimated by the width of the dip.

Extending the idea of charge transfer fluctuations, we also proposed the
forward-backward multiplicity correlation
as a possible signal for the presence of a QGP.
A case study with an embedded QGP component
in a few hadronic event generators is under way.

\appendix
\section{Effect of net positive charges \label{app::netcharge}}

For the uncorrelated charged particles in the system, keeping the total
charge
constant in the system, we find the charge transfer fluctuations to be
  \begin{eqnarray}
    \bar{D}_u(y) &=& \frac{\langle M_{\rm ch} \rangle}{4}
     \int_{-y_{\rm o} }^{ y_{\rm o} }{dy' g(y')  }
    \nonumber \\
    &&- \frac{\langle M_{\rm ch} \rangle}{4}
      \left( \int_{y}^{y_{\rm o}} {dy' g(y') }
           - \int_{-y_{\rm o}}^{y}{dy' g(y') } \right)^2
    \, .
  \nonumber \\
  \label{eq:app:ChTrans-UnpairedCharges}
  \end{eqnarray}
In deriving the above result, We have assumed that the positive and
negative
charges have the same normalized distribution, $g(y)$. Otherwise, we have
to
count the contributions from both positive and negative charges separately
in
Eq.(\ref{eq:app:ChTrans-UnpairedCharges}), and additionally there will be
an
extra term corresponding to the difference of the positive and negative
charge
forward-backward asymmetry.

The first term in Eq.(\ref{eq:app:ChTrans-UnpairedCharges}) comes from the
overall fluctuations of charged particle number in the observation window
and
its value is the same as in the net charge
fluctuations case (except a factor of $4$). The second term is due to the
forward-backward asymmetry nature in the charge transfer definition. In
the
limit that the observation window is sufficient large,
Eq.(\ref{eq:app:ChTrans-UnpairedCharges}) reduces to the second integral
in
Eq.(\ref{eq:ChTr:ChTrFl-dist-func}).

In the case that the system has both uncorrelated charges and
correlated charges, we only need to add the result in
Eq.(\ref{eq:app:ChTrans-UnpairedCharges})
to the previous result for correlated charges,
Eq.(\ref{eq:OneComp:ChargeTrans-useSymmetry}).
This will not change the quantitative features of the charge transfer
fluctuations as a function of rapidity.
The first term in
Eq.(\ref{eq:app:ChTrans-UnpairedCharges}) is independent of $y$
and will not affect any of our discussion except adding a constant.
The second term is a decreasing function of $y$. So, the contribution
from uncorrelated charges is still a decreasing function of
forward-backward rapidity cut $y$. This is in line with
the results for correlated
charges. The decreasing trend of the charge transfer fluctuations
as a function of the forward-backward rapidity cut $y$ in uniform system
is unchanged by this additional contribution. The existence of an
increasing
segment will still be a signal of a second phase with smaller charge
correlation length.

Since in realistic heavy-ion collisions at RHIC energies, most of
positive and negative charges are created together with an opposite
charge to conserve the net total charges, we can safely assume that
uncorrelated charges are rare.
The net positively charged particles originating from the projectile
and the targets is only a small fraction of the total number of
charged particles in RHIC energy heavy-ion collisions.
For this reason, the corrections from uncorrelated
charges are ignored in the most of this study. The qualitative
features of the charge transfer fluctuations will not be sensitive
to this correction term.

We can make a simple estimate of the corrections
from these uncorrelated charges
assuming the uncorrelated charges are from
the protons in the initial collision
system.  The maximum of the second
term in Eq.(\ref{eq:app:ChTrans-UnpairedCharges}) scales as $p^2 M_+ / M_0
$,
where $p$ is the fraction of observed uncorrelated
charges to the total number of uncorrelated charges. In RHIC energy
heavy-ion
reactions, $M_+ / M_0 \sim 0.04$ and $p$ is typically around $5\%$ in the
central
region. Indeed, the corrections from uncorrelated charges
is quite small, of order $10^{-4}$. The
correction to the net charge fluctuations from the uncorrelated charges
are on
the same order of magnitude as the corrections to the charge transfer
fluctuations. The net charge fluctuations
$D_c(y_{\rm o})$ and the total number of
charges $N_{\rm ch}(y_{\rm o})$ both acquires additional
terms and they are both equal
to $4$ times the first term in Eq.(\ref{eq:app:ChTrans-UnpairedCharges}).

\section{Nonzero Charge Transfer Case  \label{app::Asym}}
When the charge transfer $u(y)$ in Eq.(\ref{eq:ChTr:fbcharge-asymmetry})
does
not average to zero, the charge transfer fluctuations will acquire
additional
terms that are quadratic to the average charge transfer.

In the neutral cluster model, the full result for the charge transfer
fluctuations is:
  \begin{eqnarray}
    \bar{D}_u(y) &=& \frac{\langle M_0 \rangle}{2}
    (W_L + W_R + 2 W_y)
    \nonumber \\
   &+&  \frac{\langle u(y) \rangle^2}{\langle M_0 \rangle^2}
     \left( \langle \delta M_0^2 \rangle - \langle M_0 \rangle \right)
  \label{eq:app:ChTrans-all-pairedCharges}
  \end{eqnarray}
The weights for left and right edges observation window $(-y_{\rm
o},y_{\rm
o})$ and for the forward-backward rapidity cut $y$ are defined as:
  \begin{eqnarray}
  W_L &=&  \int_{-\infty}^{-y_{\rm o}} dy' \int_{-y_{\rm o}}^{y_{\rm
o}}dy''
  \, f_0(y',y'')
  \nonumber \\
  W_R &=&  \int_{-y_{\rm o}}^{y_{\rm o}} dy' \int_{y_{\rm o}}^{\infty}
dy''
  \, f_0(y',y'')
  \nonumber \\
  W_y &=& \int^y_{-y_{\rm o}} dy'\int_y^{y_{\rm o}} dy'' \, f_0(y', y'')
   \, .
  \label{eq:app:ChTrans-all-weights}
  \end{eqnarray}
The last term in Eq.(\ref{eq:app:ChTrans-all-pairedCharges}) stems from
the
nonzero average charge transfer in the system. An estimate would give
$\langle
u(y) \rangle \leq 10$ and
$\langle \delta M_0^2 \rangle - \langle M_0 \rangle \sim 0.1 M_0$,
and the error from
neglecting this nonzero average charge transfer is typically less than
$10^{-5}$.

\section{Solution of Eq.(\ref{eq:ChTr:onlyf0})}
\label{app:solution}

The Thomas-Chao-Quigg equation for neutral cluster distribution function
is
given by
 \be
 \int_{-\infty}^z dy\int_z^\infty dx\, f_0(x, y)
 =
 \kappa \int_{-\infty}^\infty dy\, f_0(y, z)
 \ee
 We make an ansatz:
 \be
 f_0(x, y) = g(r)\, F(Y)
 \ee
 where $r = x-y$ and $Y = (x+y)/2$ and with $g(-r) = g(r)$.
 Changing variables to $r$ and $Y$, the above equation becomes
 \begin{equation}
 \int_0^\infty dr\, g(r)
 \int_{z-r/2}^{z+r/2} dY \, F(Y)
 =
 \kappa\int_{-\infty}^\infty dr\, g(r)\, F(z + r/2)
 \end{equation}
 We can now Taylor-expand both
 $\int_{z-r/2}^{z+r/2} dYF(Y)$ and $F(z+r/2)$ with
 respect to $r$ and get the following relationship between the moments of
 $g(r)$
 \be
 R_{2n+1}/R_{2n}
 =
 2\,\kappa\, (2n+1)
 \ee
 where
 \be
 R_s \equiv \int_0^\infty dr\, g(r)\, r^s
 \ee
 and we used the fact that $g(r)$ is an even function.

 Note that
 \be
 \int_0^\infty dx\, e^{-x/2\kappa}\, x^n = 2^{n+1} \kappa^{n+1} n!
 \ee
 so that
 \be
 {\int_0^\infty dx\, e^{-x/2\kappa}\, x^{2n+1}
 \over
 \int_0^\infty dx\, e^{-x/2\kappa}\, x^{2n} }
 & = &
 {2^{2n+2} \kappa^{2n+2} (2n+1)!
 \over
 2^{2n+1} \kappa^{2n+1} (2n)!  }
 \non
 & = &
 2\kappa (2n+1)
 \ee
 Therefore
 \be
 g(r) = C\, \exp(-|r|/2\kappa)
 \ee
 where $C$ is a normalization constant.
 Hence, the solution of Eq.(\ref{eq:ChTr:onlyf0}) is given by
 \be
 f_0(x, y) = {\cal N}\, \exp(-|x-y|/2\kappa)\, F( (x+y)/2 )
 \ee
 where $F(Y)$ can be quite arbitrary as long as its derivatives are all
 finite and the integrals in Eq.(\ref{eq:ChTr:onlyf0}) are well defined.

\begin{acknowledgments}
The authors thank C.Gale, V.Topor Pop,
V.Koch and G.Westfall for stimulating discussions and J.Barrette
for his critical reading of the manuscript.
S.J.~is supported in part by the Natural Sciences and
Engineering Research Council of Canada and by le Fonds
Nature et Technologies of Qu\'ebec.
S.J.~also
thanks RIKEN BNL Center and U.S. Department of Energy [DE-AC02-98CH10886]
for
providing facilities essential for the completion of this work.
\end{acknowledgments}

\end{document}